\documentclass{revtex4}

\usepackage{graphicx}
\usepackage{amsmath,amssymb}

\begin{document}

\title{Pair Partitioning in time reversal acoustics}

\author{Hern\'an L. Calvo}
\email{hcalvo@famaf.unc.edu.ar}
\affiliation{Facultad de Matem\'atica, Astronom\'ia y F\'isica, Universidad Nacional de C\'ordoba, Ciudad Universitaria, 5000 C\'ordoba, Argentina}

\author{Horacio M. Pastawski}
\affiliation{Facultad de Matem\'atica, Astronom\'ia y F\'isica, Universidad Nacional de C\'ordoba, Ciudad Universitaria, 5000 C\'ordoba, Argentina}

\begin{abstract}
Time reversal of acoustic waves can be achieved efficiently by the persistent control of excitations in a finite region of the system. The
procedure, called Time Reversal Mirror, is stable against the inhomogeneities of the medium and it has numerous applications in medical 
physics, oceanography and communications. As a first step in the study of this robustness, we apply the Perfect Inverse Filter procedure that
accounts for the memory effects of the system. In the numerical evaluation of such procedures we developed the Pair Partitioning method for a
system of coupled oscillators. The algorithm, inspired in the Trotter strategy for quantum dynamics, obtains the dynamic for a chain of 
coupled harmonic oscillators by the separation of the system in pairs and applying a stroboscopic sequence that alternates the evolution of
each pair. We analyze here the formal basis of the method and discuss his extension for including energy dissipation inside the medium.
\end{abstract}

\maketitle

\section{Introduction}

In the recent years, the group of M. Fink developed an experimental technique called Time Reversal Mirror (TRM) \cite{fink99} that allows
time reversal of ultrasonic waves. An ultrasonic pulse is emitted from the inside of the control region (called cavity) and the excitation 
is detected as it escapes through a set of transducers placed at the boundaries. These transducers can act alternatively like microphones or
loudspeakers and the registered signal is played back in the time reversed sequence. Thus, the signal focalizes in space and time in the 
source point forming a Loschmidt Echo \cite{jalabert01}. It is remarkable that the quality of focalization gets better by increasing the 
inhomogeneities inside the cavity. This property allows for applications in many fields \cite{fink03,edelman02}. In order to have a first 
formal description for the exact reversion, we introduced a time reversal procedure denoted Perfect Inverse Filter (PIF) in the quantum 
domain \cite{pastawski07}. The PIF is based in the injection of a wave function that precisely compensates the feedback effects by means of 
the renormalization of the registered signal in the frequency domain. This also accounts for the correlations between the transducers. 
Recently, we proved that these concepts apply for classical waves \cite{calvo07}. We applied it to the numerical evaluation of the reversal 
of excitations in a linear chain of classical coupled harmonic oscillators with satisfactory results. A key issue in assessing the stability 
of the reversal procedure is have a numerical integrator that is stable and perfectly reversible. Therefore, we developed a numerical 
algorithm, the Pair Partitioning method (PP), that allowed the precise test of the reversal procedure. In the next section we develop the 
main idea of the method and then we use it to obtain the numerical results that we compare with the analytical solution for the homogeneous 
system. Additionally, we introduce a method to approximate the solution of an infinite system using a finite one. For this we introduce a
non-homogeneous fictitious friction term that can simulate the diffusion of the excitation occurring in an unbounded system. These strategies
are tested through a numerical simulation of a time reversal experiment.

\section{Wave dynamics in the Pair Partitioning method}

\begin{figure}
\includegraphics{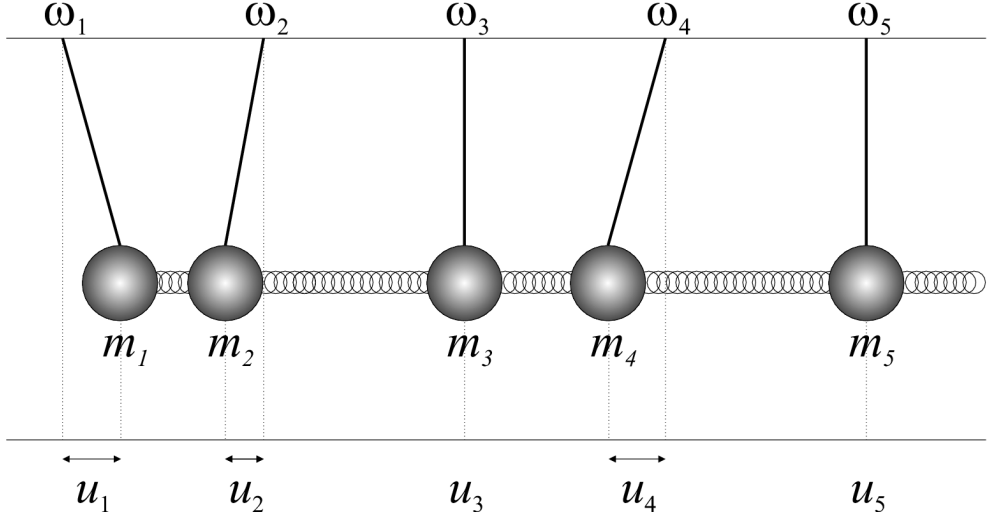}
\caption{Scheme of the $N$ coupled oscillators system to be solved by the PP numerical method.}
\label{fig_scheme}
\end{figure}

The system to be used is shown in the figure \ref{fig_scheme}: a one-dimensional chain of $N=2s$ coupled oscillators with masses $m_{i}$ and
natural frequencies $\omega _{i}$ that can be represented as a set of coupled pendulums.

If $p_{i}$ denotes the impulse and $u_{i}$ the displacement amplitude from the equilibrium position for the $i$th oscillator, the Hamiltonian
writes%

\begin{equation}
\mathcal{H}=\sum_{i=1}^N\left(\frac{p_i^2}{2m_i}+\frac{m_i\omega_i^2}{2}u_i^2\right)+\sum_{i=1}^{N-1}\frac{K_i}{2}\left(u_{i+1}-u_i\right)^2,
\end{equation}%

where $K_i$ is the elastic coefficient that accounts for the coupling between the oscillators $i$ and $i+1$. Notice that we could rewrite the
Hamiltonian in terms of each coupling separating it in non-interacting terms including even pairs and odd pairs each:%

\begin{equation}
\begin{aligned}
\mathcal{H}&=\mathcal{H}_{\mathrm{odd}}+\mathcal{H}_{\mathrm{even}}=\mathcal{H}_{1,2}+\mathcal{H}_{3,4}+\ldots+\mathcal{H}_{N-1,N}\\
&+\mathcal{H}_{2,3}+\mathcal{H}_{4,5}+\ldots+\mathcal{H}_{N-2,N-1},
\end{aligned}
\end{equation}%

with%

\begin{equation}
\mathcal{H}_{n,n+1}=\frac{p_n^2}{2\tilde{m}_n}+\frac{\tilde{m}_n\tilde{\omega}_n^2}{2}u_n^2+\frac{p_{n+1}^2}{2\tilde{m}_{n+1}}+%
\frac{\tilde{m}_{n+1}\tilde{\omega}_{n+1}^2}{2}u_{n+1}^2+\frac{K_n}{2}\left(u_{n+1}-u_n\right)^2,
\end{equation}%

and%

\begin{equation}
\begin{aligned}
\tilde{m}_1 &=m_1,\ \tilde{\omega}_1=\omega_1,\ \tilde{m}_N=m_N,\ \tilde{\omega}_N=\omega _N,\\
\tilde{m}_n &=2m_n,\ \tilde{\omega}_n=\omega _n/2,\ n=2,\ldots,N-1.
\end{aligned}
\end{equation}%

A good approximation to the overall dynamics, inspired in the Trotter method used in quantum mechanics \cite{deraedt96}, can be obtained 
solving analytically the equations of motion for each independent Hamiltonian in a time step $\tau$. Therefore, the pair 
$\mathcal{H}_{n,n+1}$ has%

\begin{equation}
\begin{aligned}
\ddot{u}_{n} &=-\left(\tilde{\omega}_n^2+\frac{K_n}{\tilde{m}_n}\right)u_n+\frac{K_n}{\tilde{m}_n}u_{n+1},\\
\ddot{u}_{n+1} &=-\left(\tilde{\omega}_{n+1}^2+\frac{K_n}{\tilde{m}_{n+1}}\right)u_{n+1}+\frac{K_n}{\tilde{m}_{n+1}}u_n.
\end{aligned}
\end{equation}%

At each small time step $\tau$, the evolution for the even couplings is obtained and the resulting positions and velocities are used as 
initial conditions for the set of Hamiltonians accounting for odd couplings and so on. Since the equations of motion are solved separately, 
we could consider only the two coupled oscillators system, e.g.%

\begin{equation} \label{eq_mov}
\begin{aligned}
\ddot{u}_1 &=-\omega_1^2 u_1+\omega_{12}^2 u_2,\\
\ddot{u}_2 &=-\omega_2^2 u_2+\omega_{21}^2 u_1.
\end{aligned}
\end{equation}%

For this system, it is easy to obtain the corresponding normal modes%

\begin{equation} \label{eq_normal}
U_{\pm}=\mp\frac{\omega_{21}^2}{\omega_{+}^2-\omega_{-}^2}u_{1}\pm\frac{\omega_1^2-\omega_{\mp}^2}{\omega_{+}^2-\omega_{-}^2}u_2,  
\end{equation}%

with characteristic frequencies%

\begin{equation}
\omega _{\pm}^2=\frac{\omega_1^2+\omega_2^2}{2}\pm\sqrt{\left(\frac{\omega_1^2-\omega_2^2}{2}\right)^2+\omega_{12}^2\omega_{21}^2}.
\end{equation}%

From (\ref{eq_normal}), $U_{\pm}(t)$ are obtained and these values are used for the evolution after the temporal step 
$t \rightarrow t+\tau$, i.e.%

\begin{equation} \label{eq_evnorm}
\begin{aligned}
U_{\pm}(t+\tau) &=U_{\pm}(t)\cos(\omega_{\pm}\tau)+\dot{U}_{\pm}(t)\sin(\omega_{\pm}\tau)/\omega_{\pm},\\
\dot{U}_{\pm }(t+\tau ) &=\dot{U}_{\pm}(t)\cos(\omega_{\pm}\tau)-U_{\pm}(t)\sin(\omega_{\pm}\tau)\omega_{\pm}.
\end{aligned}
\end{equation}%

Once we have $U_{\pm}(t+\tau )$ and $\dot{U}_{\pm}(t+\tau )$ we can go back to the natural basis by means of the inverse of 
(\ref{eq_normal})%

\begin{equation} \label{eq_inv}
\begin{aligned}
u_1 &=\frac{\omega_1^2-\omega _{+}^2}{\omega_{21}^2}U_{+}+\frac{\omega_1^2-\omega_{-}^2}{\omega_{21}^2}U_{-},\\
u_2 &=U_{+}+U_{-}.
\end{aligned}
\end{equation}%

Then, one obtains the displacements and momenta for all oscillators at time $t+\tau$. 
The above steps are summarized in the Pair Partitioning algorithm:

\begin{enumerate}
\item Determine all the masses and natural frequencies of the partitioned system $\tilde{m}_{n},\tilde{\omega}_n$.

\item For even couplings $\mathcal{H}_{2,3},\mathcal{H}_{4,5},\ldots$, rewrite the initial conditions $\left\{u_{i}(0),\dot{u}_{i}(0)\right\}$
for the normal modes according to (\ref{eq_normal}).

\item Calculate the normal modes evolution for even couplings, according to (\ref{eq_evnorm}) and obtain $\left\{u_i^{\mathrm{even}}(\tau),\dot{u}_i^{\mathrm{even}}(\tau)\right\}$ from (\ref{eq_inv}).

\item Calculate the normal modes for odd couplings $\mathcal{H}_{1,2},\mathcal{H}_{3,4},\ldots$, using the recent positions and velocities.

\item Calculate the normal modes evolution for odd couplings and give the positions and velocities $\left\{u_i(\tau),\dot{u}_i(\tau)\right\}$.

\item Go back to the step 2 with $\left\{u_i(\tau),\dot{u}_i(\tau)\right\}$.
\end{enumerate}

\begin{figure}
\includegraphics{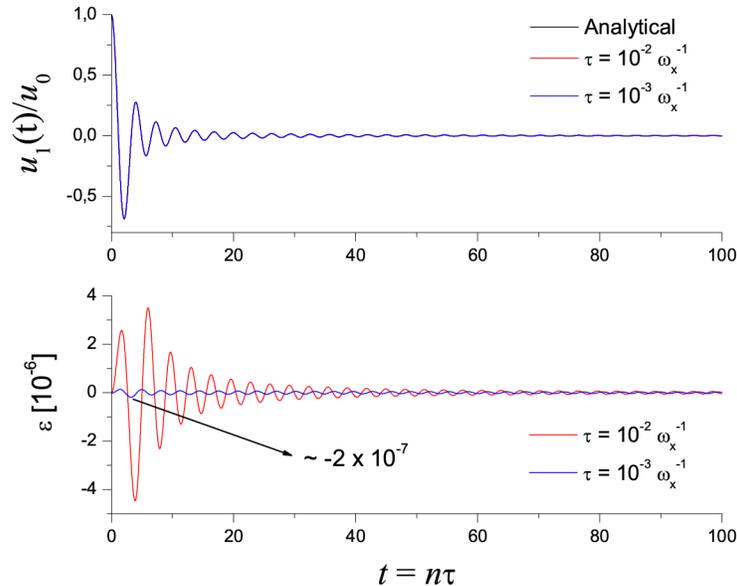}
\caption{Time evolution for the surface oscillator. Top, comparison between the
analytical and numerical results shows all curves superposed. Bottom, error in the displacement for both
temporal steps.}
\label{fig_ordered}
\end{figure}

Therefore, applying $n$ times the PP algorithm we obtain the positions and velocities for all oscillators at time $t=n\tau$.\\
As an example we consider the homogeneous system where the $N$ oscillators have identical masses and the only natural frequencies correspond
to the surface $\omega_1=\omega_N=\omega_{\mathrm{x}}=\sqrt{K/m}$. The displacement amplitude of the $i$th oscillator due to an initial
displacement in the $j$th oscillator can be expressed analytically as%

\begin{equation}
u_{i \leftarrow j}^{\mathrm{th.}}(t)=\frac{2}{N+1}\sum_{k=1}^N\sin\left(i\frac{k\pi}{N+1}\right)\sin\left(j\frac{\pi}{N+1}\right)\cos(%
\omega_kt),
\end{equation}%

with%

\begin{equation}
\omega_k=\omega_{\mathrm{x}}\sin\left(\frac{k\pi}{2(N+1)}\right),
\end{equation}%

the characteristic frequency for the $k$th normal mode. In the figure \ref{fig_ordered} the analytical and numerical results are compared for
the surface oscillator displacement in a case when all the oscillators were initially in their equilibrium positions except $u_1(0)=u_1^{0}$.
We use $N=200$ in two cases where the temporal steps are $\tau=10^{-2}\omega_{\mathrm{x}}^{-1}$ and $\tau=10^{-3}\omega_{\mathrm{x}}^{-1}$. 

We have taken $N$ such that no mesoscopics echoes \cite{pastawski95} appear in the interval of time shown. We also notice that the error%

\begin{equation}
\varepsilon(t)=\frac{\left\vert u_1^{\mathrm{th.}}(t)-u_{1}^{\mathrm{PP}}(t)\right\vert}{u_{0}},
\end{equation}%

drops as the temporal step $\tau$ diminishes. We observe a quadratic dependence $\mathrm{\max}{\varepsilon(t)}=\alpha\tau^2$ in complete
analogy with the Trotter method. In the particular case of the homogeneous system we have $\alpha\simeq 0.0445$.

\subsection{Unbounded systems as damped oscillations}

The solution of wave dynamics in infinite media remains as a delicate problem. In such case, the initially localized excitation spreads through
the systems in a way that resembles actual dissipation. In contrast, finite systems present periodic revivals, the mesoscopic echoes, that 
show that energy remains in the system. In order to get rid the mesoscopic echoes and obtaining a form of \textquotedblleft 
dissipation\textquotedblright\ using a finite number of oscillators, we add a fictitious \textquotedblleft friction\textquotedblright\ term. 
The friction coefficients $\eta _i\geq 0$ can be included between the $2$th and $3$th steps of the PP algorithm supposing that the 
displacement amplitude decays exponentially%

\begin{equation}
u_i(t)\rightarrow u_i(t)\exp(-\eta_i\tau),
\end{equation}%

as occurs in a damped oscillator in the limit $\eta /\omega \ll 1$. For the homogeneous system with $N=100$ oscillators were the cavity 
ending at site $x_R$ we choose a progressive increase in the damping as%

\begin{equation}
\eta_i=0.1\frac{i-x_R}{N-x_R},\ i=x_R,\ldots,N.
\end{equation}%

\begin{figure}
\includegraphics{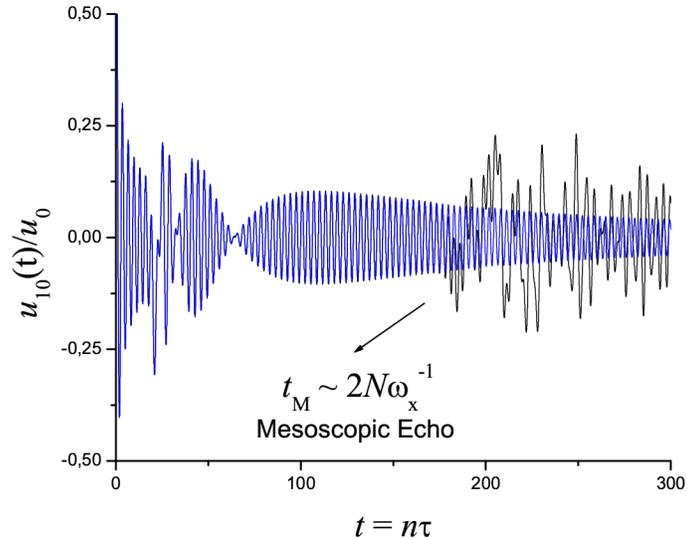}
\caption{Comparison between the undamped evolution (black) and the damped
evolution (blue) for the displacement amplitude $u_{R}(t)$.}
\label{fig_mesosc}
\end{figure}

We compare the result of this approximation with the undamped case for the displacement amplitude in $x_R=10$. The figure \ref{fig_mesosc} 
shows how the dynamics in the damped system has no mesoscopic echoes whereas in the undamped system we observe the echo at 
$t_M\simeq 2N\omega_{\mathrm{x}}^{-1}$. 

As we will see for TRM and PIF procedures, this last result is very usefull since it allows to obtain the dynamics of an open system with a 
small number of oscillators (e.g. $N\simeq 100$).

\section{Numerical test for the Perfect Inverse Filter procedure}

\begin{figure}
\includegraphics{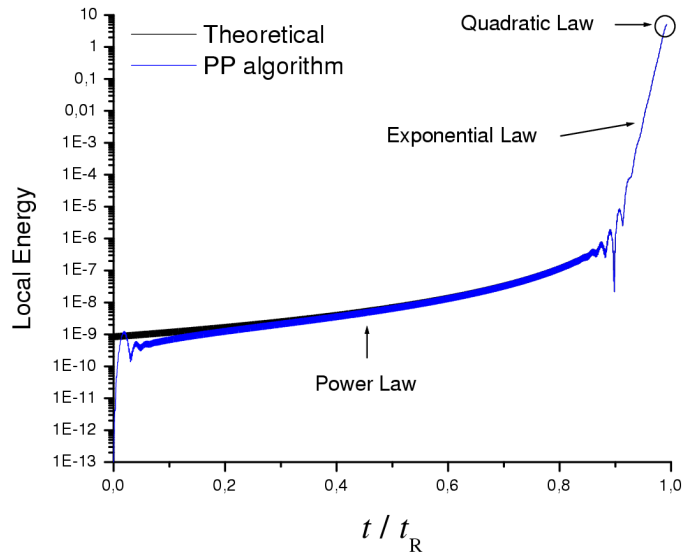}
\caption{Local energy of the surface oscillator recovering. Here, we choose for the detection time $t_R=1000\tau$.
evolution (blue) for the displacement amplitude $u_{R}(t)$.}
\label{fig_recovering}
\end{figure}

As we mentioned above, the Time Reversal Mirror procedure consists in the injection, at the boundaries of the cavity, of a signal 
proportional to that recorded during the forward propagation. In contrast, the Perfect Inverse Filter corrects this recorded signal to 
accounts for the contributions of multiple reflections and normal dispersion of the previously injected signal in a manner that their 
instantaneous total sum coincides precisely with the time reversed signal at the boundaries. The continuity of the wave equation ensures that
perfect time reversal occurs at every point inside the cavity. The procedure for such correction is described somewhere else 
\cite{pastawski07,calvo07}. Here, it is enough to notice that the imposition of an appropriate wave amplitude at the boundaries should give
the perfect reversal of an excitation originally localized inside the cavity. An example of such situation would be a 
\textquotedblleft\ surface pendulum\textquotedblright\ coupled to a semi-infinite linear chain of harmonically coupled masses \cite{calvo06}.
In such a case, we know that the energy decays in a approximately exponential way, where the decay rate can be assimilated to a 
\textquotedblleft friction coefficient \textquotedblright. However, for very short times, the local energy decays with a quadratic law 
while for very long times the exponential decay gives rise to a power law characteristic of the slow diffusion of the energy \cite{rufeil06}.
In the figure \ref{fig_recovering} we show how this overall decay is reversed by controling the amplitude at site $x_R=10$.

As long as we have been able to wait until a neglegible amount of energy is left in the cavity, the control of the boundaries is enough to 
reverse the whole dynamics inside the cavity. As a comparison, the theoretical reversal of the decay of a surface excitation in a 
semi-infinite chain is shown. We see that the region they differ is when the injected signal is still negligible as compared to the energy 
still remaining in the cavity (as a concequence of the very slow power law decay).

\section{Discussion}

We have presented a numerical strategy for the solution of the wave equation that is completely time reversible. This involves the iterative
application of the exact evolution of pairs of coupled effective oscillators where the energy is conserved, hence deserving the name of Pair
Partitioning method. This is complemented with an original strategy for dealing with wave propagation through infinite media. While various 
tests of these procedures remain to be done, we have shown, through the solution of simple but highly non-trivial examples, that the method
is numerically stable and can be used to revert wave dynamics up to a desired precision.

\end{document}